\documentclass[12pt]{article}

\newcommand{\lsim}{\raisebox{-0.5ex}{$\stackrel{<}{\sim}$}}
\usepackage{epsfig}
\usepackage{citesort}
\usepackage{overcite}
\usepackage{amssymb}

\begin{document}

\renewcommand{\abstractname}{{\normalfont ABSTRACT}}
\parskip = 12pt
\parindent = 0pt
\baselineskip= 13 pt
\setlength{\textwidth=16cm}
\setlength{\textheight=21cm}

\titlepage{
  \begin{center}
    {\bf{\normalsize{Model studies of diffusion in glassy and polymer ion conductors}}}
  \end{center}
\vspace{-0.7cm}
\begin{center}
O. D\"{u}rr${}^{1}$, P. Pendzig${}^{1}$, W. Dieterich${}^{1}$ and
A. Nitzan${}^{2}$
\end{center}
\vspace{-0.7cm}
\begin{center}
\vspace{-0.4cm}
{{\footnotesize{${}^1$Fakult\"{a}t f\"{u}r Physik,
Universi\"{a}t Konstanz, D-78457 Konstanz, Germany\\
${}^2$School of Chemistry, The Sackler Faculty of Science, Tel Aviv University,\\
Tel Aviv 69978, Israel}}}
\end{center}

\vspace{-0.9cm}
\begin{abstract}
\vspace{-0.65cm}
Ion--conducting glasses and polymer systems show several characteristic
peculiarities in their composition--dependent diffusion properties and in
their dynamic response. First we give a brief review of the current
understanding of the ion dynamics in network glasses in terms of stochastic
theories. Secondly, a model for PEO(polyethylene-oxide)--based polymer electrolytes is
described. The equation of state is calculated from the quasichemical
approximation. Using this information together with Monte Carlo simulations of
diffusion at constant volume, we obtain constant--pressure results for polymer
chain and ion diffusion as a function of ion concentration. This theory allows
us to make comparison with experiments which are carried out at constant
pressure.
\end{abstract}

\vspace{-0.6cm}
\section*{\normalsize{\bf 1. INTRODUCTION}}
\vspace{-0.4cm}

Ion conducting glassy \cite{Ang90,Ing87,Sha96} and polymeric \cite{Gra91,Rat88} materials show a large variability
in their chemical composition, local structure, ion diffusion properties and
in many other physical properties \cite{Las89}. Investigation of these amorphous systems
and of the relevant ion transport mechanisms thus constitutes a wide and
active area of research.

Generally, ion migration in these materials is connected with slow relaxation
processes on time--scales much larger than microscopic times. A well-known
example occurs in glasses, whose low--frequency dynamic response appears to
slow down continuously as temperature is lowered \cite{Ro97,Now98}. A quite different
situation is known for polymer electrolytes, where ions are strongly coupled to
the polymer network. Therefore, ion diffusion slows down and gets almost
suppressed when approaching the glass transition temperature \cite{Bru93,Fan94}. Theories,
which aim at a description of ionic motions on such extended time--scales,
will have to eliminate all fast (e.g. vibrational) degrees of freedom from the
outset. One then arrives at coarse--grained models based on stochastic ionic
moves in an environment that reflects the characteristic local structure of
the system.

Adopting this point of view, we shall first focus on several transport
anomalies in glasses and discuss the respective state of stochastic modelling
(section 2). In section 3 we introduce a lattice model for polymer
electrolytes which emphasises the mutual influence of ion and polymer chain
dynamics \cite{Ole94,Pen98}. Monte Carlo simulations of diffusion in an (NVT)--ensemble will
be supplemented by calculations of the equation of state via the quasichemical
approximation (QCA) in the version of Guggenheim \cite{Gug52} and Barker \cite{Bar52}. Combination
of both methods will enable us to obtain transport properties in an
(NpT)--ensemble. Consideration of a constant pressure $p$ is important when
comparing theory with experimental trends under varying composition, such as
the increase in the glass transition temperature or the decrease in ionic
mobilities upon adding salt \cite{Mc92,Fan94}.

\vspace{-0.6cm}
\section*{\normalsize{\bf 2. ION TRANSPORT ANOMALIES IN GLASSES}}
\vspace{-0.4cm}
Alkali--doped network glasses (oxide- or sulphide--glasses, like the systems
$M_{2}O-SiO_{2}$ or $M_{2}S-SiS_{2}$, where $M$ represents the alkali
ion) exhibit several peculiar features in their diffusion properties:
\begin{itemize}
\item[{i)}]
The activation energy $E(c)$ for dc--transport markedly decreases with
increasing alkali concentration $c$ for dilute systems (small $c$), while it
saturates for strongly doped samples. This behaviour results in a very
sensitive $c$--dependence of the ambient--temperature dc--conductivity \cite{Ing87}.
\item[{ii)}]
The Haven--ratio $H(c)$ shows a steep decrease with $c$ and also saturates for
larger $c$ \cite{Kel80}.
\item[{iii)}]
Conductivity dispersion is an important effect where, apparently, different
mechanisms are active. At lower frequencies, Jonschers' 'universal
ac--response' applies (power--laws in frequency, with exponents
$n_{\sigma}\simeq 0.6$ to $0.7$ \cite{Jon83,Nga96}), while at high frequencies one observes a
'nearly constant--loss'--type response, characterized by an exponent
$n_{1}\simeq 1$ \cite{Now98,Hsi96}. When temperature gets lowered, the latter process dominates
the conductivity spectra.
\item[{iv)}]
Dynamical processes which govern the ac--conductivity spectra also manifest
themselves in quasi--elastic neutron scattering, ultrasonic attenuation and
nuclear--spin relaxation \cite{Fu93}.
\item[{v)}]
Mixed--alkali systems, e.g. $\mathrm (Na_{2}O)_{x}(K_{2}O)_{1-x}$--$\mathrm SiO_{2}$, show a
highly nonlinear dependence of $\mathrm Na-$ and $\mathrm K$--diffusion
constants and of other physical properties on the ratio of mixing $x$ \cite{Day76}.
\end{itemize}

Up to now, no theory exists that accounts for all of these
features. Nevertheless, deeper insight into their microscopic origin has
emerged from specific model studies. Here, we only briefly address some of the
most recent approaches and refer the interested reader to the original
literature.

Properties of dispersive transport and composition--dependent effects in
single--alkali glasses can be understood with reasonable accuracy from
stochastic lattice gas models based on structural disorder and
Coulomb--interactions \cite{Ma91,Pet92,Ma95}. In particular, the 'counter--ion model' has proved
successful in explaining the behaviour of weakly doped glasses \cite{Kno96}. It also
proposes a mechanism for the 'nearly constant loss' response in terms of
correlated, dipolar reorientation processes \cite{Pe98}. On the other hand, for
explaining mixed--alkali--effects, a quite different approach in terms of the
'dynamic structure model' has been suggested, which involves the concept of
lattice mismatch and a dynamical adjustment of the local structure to each ionic
species \cite{Ma92,Bu94}. The design of a unique model by a proper combination of these
ideas remains a task of future research.

\vspace{-0.6cm}
\section*{\normalsize{\bf 3. LATTICE MODEL OF POLYMER IONIC CONDUCTORS}}
\vspace{-0.4cm}
As pointed out in the Introduction, ionic and network degrees of freedom in
polymer ionic conductors are strongly coupled and influence each other in a
fundamental way. For example, in PEO (polyethylene--oxide)--based
electrolytes, binding of cations to the electro--negative oxygen atoms in the
chain molecules can introduce crosslinks into the polymer network. Local
fluctuations of chains, which are essential for providing open pathways for
the ions, therefore get reduced and ionic mobilities decrease. At the same
time the network viscosity rises. Clearly, with increasing ion concentration
$c$ these effects get enhanced, but they are expected to saturate when the
number of cations approaches the number of binding sites. Some of these
aspects have been studied recently with the aid of stochastic models
consisting of an interacting system of lattice chains and two species of
point particles, representing cations and anions, respectively. Details of
this model and the simulation technique are described in Ref.\cite{Pen98}. Our aim here
is to investigate a somewhat simplified model which contains only one species
of point--particles ("ions") which can bind to the chains. This
simplification, of course, ignores all features connected with anion
diffusion, yet it allows us to investigate the profound influence of ions on
the polymer network rigidity. In addition, it greatly facilitates the problem
to convert results for diffusion at constant volume to the corresponding
results for constant pressure. This will enable us to describe experiments on
polymer electrolytes under varying salt content, which normally are carried
out under constant pressure.

The model we consider involves lattice polymer chains (on a simple cubic
lattice of spacing $a$) which consist of two types of beads, C--beads which do
not interact with the point--particles and X--beads which attract
point--particles with strength $\epsilon$. In analogy to PEO the sequence of
beads is taken as $C(XCC)_{n}$, so that the length of chains is given by
$r=3n+1$. Throughout this work we assume $r=13$. All the beads repel each
other with a common strength, which again is taken as $\epsilon$. In our
simulations, see section 5, the equilibration procedure and the rules for
elementary chain and ion motions are taken as in Ref. \cite{Pen98}. 

\vspace{-0.6cm}
\section*{\normalsize {\bf 4. EQUATION OF STATE}}
\vspace{-0.4cm}
To extract the pressure from Monte Carlo runs for a lattice system normally is
a difficult task, although several methods are known in principle \cite{Dick87,Mac95,Pen97}. One
reason is the problem of thermalization following a stochastic volume
change. Therefore, with respect to the equation of state of our model, we
shall rely on the quasi--chemical approximation (QCA), which requires solving
a set of nonlinear equations for the quantities $N_{ij}$, which denote the number
of nearest neighbour pairs consisting of molecules of type $i$ and $j$. The
version by Barker \cite{Bar52}, which we adopt here, allows us to deal with
heterogeneous chains as described in the foregoing section \cite{Die98}.

Rather than going into details, we merely give the final result for the
pressure $p$, obtained from

\begin{equation}\label{P}
-\frac{pa^{3}}{k_{B}T}=\ln \left( \frac{N_{0}}{M} \right) +\frac{z}{2}\ln
\left( \frac{2N_{00}M}{zN_{0}^{2}}\right)
\end{equation}

Here $M$, $N_{0}$ and $N_{00}$ denote the numbers of lattice sites, vacant
sites and vacancy pairs, respectively, to be calculated from the
QCA--equations as a function of the numbers of molecules and temperature.

Isobars for different numbers of ions are shown in Fig. 1 at a fixed (reduced)
pressure
$pa^{3}/\epsilon=0.35$. When we add ions at a high temperature, the system
swells, but the amount of swelling is less than what one would expect for
non--interacting point particles. This, clearly, is a consequence of the
attractive ion--polymer interaction, which even leads to a reversed trend at
low temperatures (see Fig. 1). Let us remark that inclusion of anions in an extended model
would favour a volume increase.

\begin{figure}[htb]
  \begin{center}
    \epsfig{file=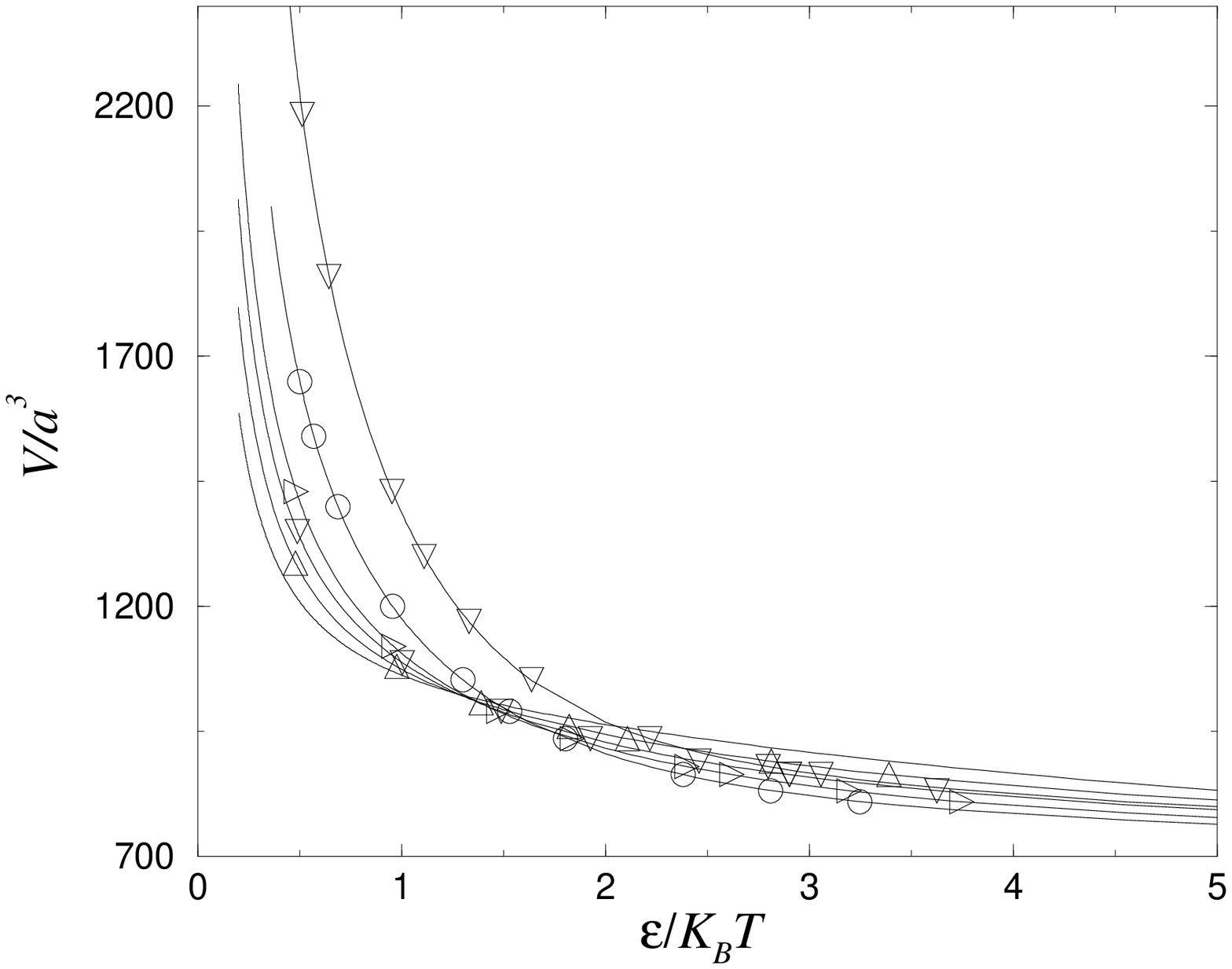,width=0.7\linewidth,angle=0} 
  \end{center}
   \footnotesize{Figure 1: QCA--isobars $(pa^{3}/\epsilon=0.35)$ for different
   concentrations of point particles. Values for volume and temperature
   indicated by special points are used in the simulations for $D^{(P)}$
   and $D^{(I)}$ shown in Fig. 2. At high temperatures ($\epsilon/k_{B}T
   \lesssim 1$) the different curves refer to (from below):
   $N_{1}=0,20,40,60,120,240$.}
  \leavevmode
\end{figure}

\vspace{-0.6cm}
\section*{\normalsize {\bf 5. DIFFUSION PROPERTIES UNDER CONSTANT PRESSURE}}
\vspace{-0.4cm}
Next we turn to Monte Carlo simulations of the long--time diffusion constants
$D^{(I)}$ and $D^{(P)}$ for ions and for the centres--of--mass of polymer
chains, respectively. These quantities are deduced as usual from the
corresponding time--dependent mean--square displacements. Thereby we choose a
system with $N_{P}=31$ chains of length $r=13$ and with a varying number
$N_{1}$ of ions. Given a fixed pressure $p$  and temperature $T$, we carry
out these simulations by choosing a nearly cubic simulation box of size
$L_{1}\times L_{2}\times L_{3}=V(p,T)/a^{3}$ (with  periodic boundary
conditions) such that the volume $V(p,T)$ corresponds to that obtained from
the QCA (see Fig. 1).
Fig. 2 shows the $T$--dependence of both $D^{(I)}$ and $D^{(P)}$ at a pressure
given by $pa^{3}/\epsilon=0.35$ for various amounts of ions. As $T$ goes to
infinity, we approach the dilute limit $(V\rightarrow \infty)$, and chain
diffusion constants for different $N_{1}$ assume a common value (that
corresponds to the athermal limit). As temperature is lowered, the different
curves in Fig. 2 show a downward curvature of a form which is well represented
by the Vogel--Tammann--Fulcher (VTF) law,

\begin{figure}[htb]
   \begin{center}
    \epsfig{file=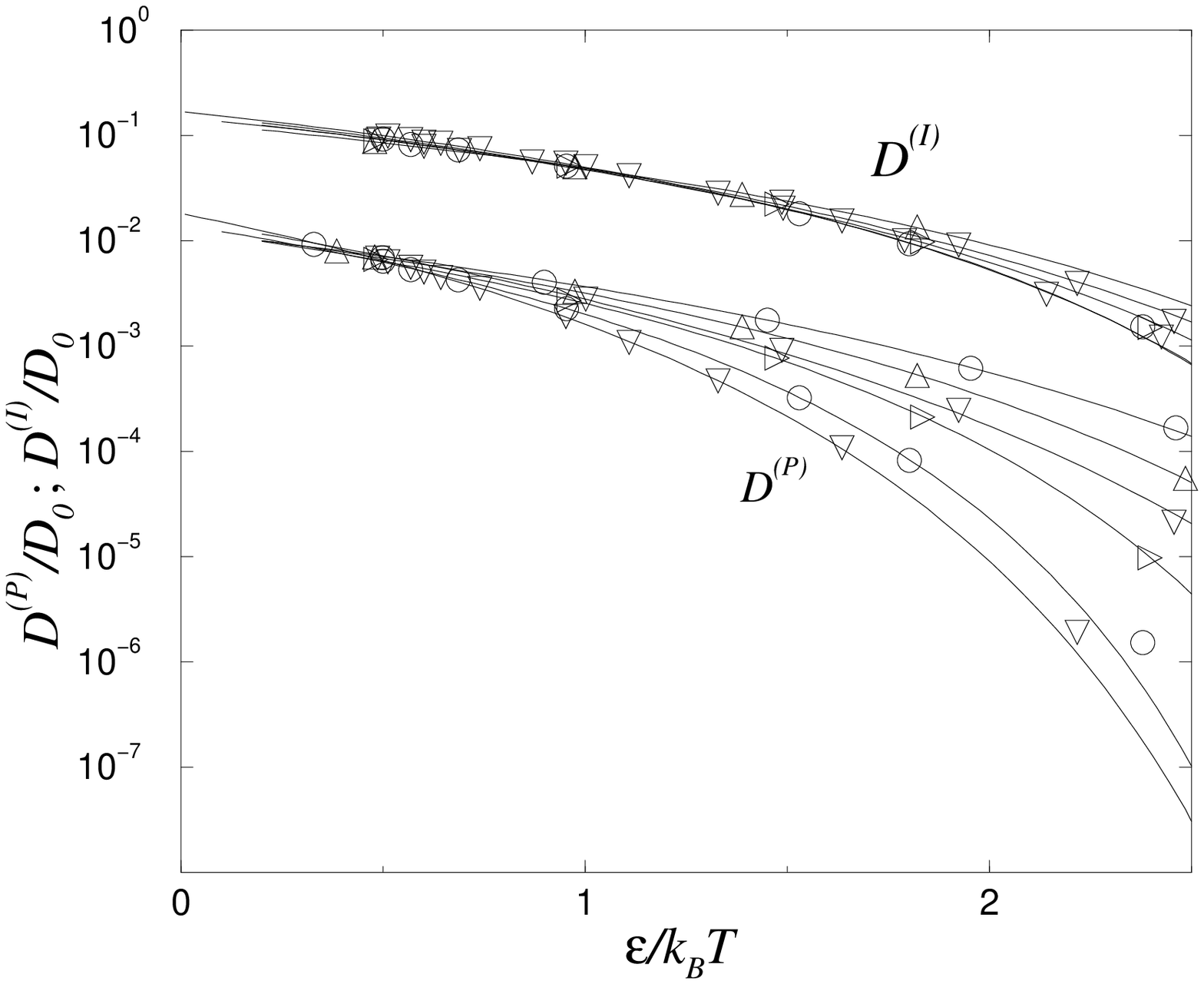,width=0.7\linewidth,angle=0} 
  \end{center}
  \footnotesize{Figure 2: Simulated polymer chain and ion diffusion constants,
     $D^{(P)}$ and $D^{(I)}$,  
     versus inverse temperature at a constant pressure,
     $pa^{3}/\epsilon=0.35$. The different sets of data and (fitted) curves
     refer to different numbers of ions (from above): $(N_{1}=0 ($O$),
     20 (\triangle), 40 (\triangledown), 60 (\triangleright), 120 ($O$), 240(\triangledown)$. At a given temperature, the sizes of simulation cells were chosen
     according to the volumes deduced from the QCA, see the special points and
     corresponding symbols in Fig. 1. Continuous lines represent fits in terms
     of the VTF--law.}
  \leavevmode
\end{figure}

\begin{equation}\label{D}
D^{(\alpha)}(T,N_{1})=D^{(\alpha)}_{\infty}\exp \left(
  -\frac{E^{\alpha}(N_{1})}{T-T^{(\alpha)}(N_{1})} \right)
\end{equation}

Here $\alpha$ stands for $I$ and $P$. The quantities $D^{(\alpha)}_{\infty}$
denote the diffusion constants at infinite temperature,
$D^{(I)}_{\infty}=D_{0}$ being the diffusion constant of a free particle, and
$D^{(P)}_{\infty}/D_{0}\simeq 10^{-2}$. $E^{\alpha}(N_{1})$ is an energetic
parameter and $T^{(\alpha)}(N_{1})$ the VTF--temperature.

A first important observation is that both $D^{(I)}$ and $D^{(P)}$ are reduced
when we increase $N_{1}$, as a result of an increased number of cations which
bind to the chains (X--beads) and possibly form crosslinks. As a consequence,
we observe rising VTF--temperatures $T^{(\alpha)}(N_{1})$, as displayed in
Fig. 3. Generally, the VTF--temperature represents a lower bound to the glass
transition temperature $T_{g}$. Our simulations with respect to the diffusion
of chains therefore are consistent with
the experimentally observed increase in $T_{g}$ with the ion
content. Interestingly, for weakly doped polymers ($N_{1}\lsim 60$),
$T^{(I)}(N_{1})$ and $T^{(P)}(N_{1})$ roughly coincide, which is indicative
of a strong coupling between ions and chains. However, as $N_{1}$ becomes
comparable with the number of X--beads, which in the present case is
$4N_{P}=124$, the influence of $N_{1}$ on the ion and chain mobilities
diminishes, as seen by comparing the data for $N_{1}=120$ and $N_{1}=240$ in
Fig. 2. Simultaneously, the VTF--temperatures saturate, as shown in Fig. 3. At
saturation, $T^{(I)}<T^{(P)}$, which means that for $T\simeq T^{(P)}$ the ions maintain a certain mobility while the
system of chains freezes. The reason is that in the strongly doped regime only
part of the ions will bind to the X--beads. These results correctly describe
the experimental trend in the glass transition temperature of PEO--based
electrolytes as a function of ion content in an appropriate regime of doping
\cite{Fan94,Mc92}. At even larger ion concentrations, however, the experimental $T_{g}$ drops
after passing a maximum. This is possibly due to ion clustering effects, not
taken into account in the present model.
\begin{figure}[htb]
  \begin{center}
    \epsfig{file=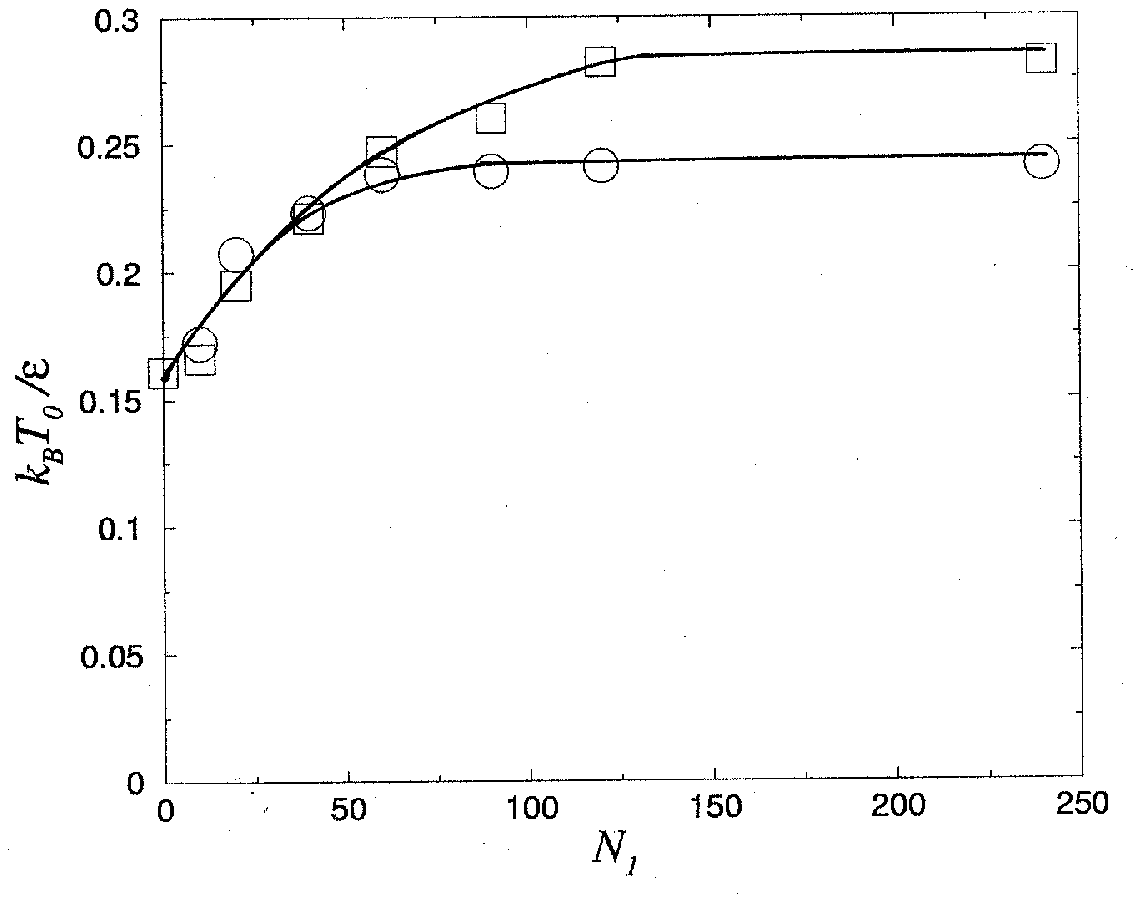,width=0.7\linewidth,angle=0} 
  \end{center}
  \footnotesize{Figure 3: VTF--temperatures for ions $(T^{(I)})$ and polymer
    chains $(T^{(P)})$ versus 
    the number of ions, as deduced from the data in Fig. 2. (Continuous lines
    are guides to the eye.)}
  \leavevmode
\end{figure}
\vspace{-0.6cm}
\section*{\normalsize {\bf 6. SUMMARY AND OUTLOOK}}
\vspace{-0.4cm}

Diffusion properties of amorphous solid electrolytes were discussed,
emphasizing the actual capabilities of semi--microscopic (stochastic) models
and Monte Carlo simulation to account for the experiments. Naturally, in this
context we were interested only in experimental key features, which are common
to many different materials.

After pointing out several characteristic ion transport properties of glasses
together with successes and persisting problems in their theoretical
understanding, we turned to a specific model for PEO--based polymer
electrolytes. Composition--dependent diffusion properties and the tendency to
undergo a glass--transition were studied under constant pressure. This was
achieved by combining constant--volume simulations with the equation of state
derived from the QCA.

The model we studied contains only one species of particles (cations) besides
the polymer chains and hence does not allow us to investigate simultaneously
the diffusion of anions. A simplified treatment of anion diffusion in an
otherwise realistic model of a polymer electrolyte might be developed with the
help of the concept of dynamic bond percolation \cite{Dru85,Nitz91,Die}.

\vspace{-0.6cm}
\section*{\normalsize{\bf Acknowledgements}}
\vspace{-0.4cm}
This work was supported in part by the Lion--Foundation.

\newpage
\bibliographystyle{alpha}

\end{document}